\documentclass[twoside,12pt]{article}
\usepackage{epsfig,graphicx}
\usepackage{wrapfig}
\usepackage{color}
\usepackage{amsmath}

\usepackage{sidecap}

\newcommand{\apjl}{ApJ}
\newcommand{\apj}{ApJ}
\newcommand{\aap}{A\&A}
\newcommand{\apjs}{ApJS}

\newcommand{\mnras}{MNRAS}
\newcommand{\aj}{AJ}
\newcommand{\nat}{Nature}

\newcommand{\be}{\begin{equation}}
\newcommand{\ee}{\end{equation}}
\newcommand{\bea}{\begin{eqnarray}}
\newcommand{\eea}{\end{eqnarray}}

\newcommand{\nuc}[2]{\ensuremath{\mathrm{^{#1}#2}}}

\newcommand{\msun}{\ensuremath{M_\odot}}

\begin{document}

\title{\vspace{1cm} Internal conversion electrons and supernova light
  curves}
\author{I.~R.\ Seitenzahl$^{1}$ \\
  $^1$Max-Planck-Institut f\"ur Astrophysik, Garching, Germany}
\maketitle
\begin{abstract}
  Radioactive decays contribute significantly to the re-heating of
  supernova ejecta. Previous works mainly considered the energy
  deposited by $\gamma$-rays and positrons produced by \nuc{56}{Ni},
  \nuc{56}{Co}, \nuc{57}{Ni}, \nuc{57}{Co}, \nuc{44}{Ti}, and
  \nuc{44}{Sc}. We point out that Auger and internal conversion
  electrons constitute an additional heat source.  At late times,
  these electrons can contribute significantly to supernova light
  curves for reasonable nucleosynthetic yields.  In particular, the
  internal conversion electrons emitted in the decay of \nuc{57}{Co}
  are an important heating channel for supernovae that have become
  largely transparent to $\gamma$-rays.  We show that when the heating
  by these electrons is accounted for, the slowing down of the light
  curves of SN~1998bw and SN~2003hv is naturally obtained for typical
  nucleosynthetic yields.  Additionally, we show that for SN 1987A the
  effects of internal conversion electrons are likely significant for
  the derivation of \nuc{44}{Ti} yields from its late time bolometric
  light curve.
\end{abstract}
\section{Introduction}
A great success of nuclear astrophysics was the demonstration that
radioactivity powers the light curves of Type Ia supernovae (SN~Ia,
most likely thermonuclear) and, at least for some events at late
times, Type Ib/c and Type II (most likely core collapse) supernovae.
Shortly after its explosion a supernova enters a state of homologous
expansion. The temperature drops during the expansion until nuclear
fusion reactions cease; radioactive decays, however, still take place,
since the matter is only slightly ionized.  It is now widely accepted
that the energy liberated in the decay chain of radioactive
\nuc{56}{Ni} is the most important nuclear source for re-heating the
supernova ejecta to temperatures high enough for the spectrum to peak
at optical wavelengths \cite{truran1967a,colgate1969a}.  At first, the
bulk of the heating is produced by the energetic $\gamma$-rays which
thermalize and deposit their energy via Compton scattering and
photoelectric absorption.  In the homologous expansion, the column
density (and therefore also approximately the Compton opacity)
decreases with time as $t^{-2}$, and the ejecta become more and more
transparent to these high energy photons. Once $\gamma$-rays escape,
the positrons produced in the decay of \nuc{56}{Co} and \nuc{44}{Sc}
were thought to be the main heating sources. Here, we draw attention
to often overlooked additional leptonic heating channels: Auger and
internal conversion electrons.  In section \ref{sec:decay} we review
the physics of nuclear decays relevant for supernova light curves. In
section \ref{sec:lc} we demonstrate the impact of internal conversion
electrons from the decay of \nuc{57}{Co} on different supernova light
curves. We conclude in section \ref{sec:conc} with an outlook how this
effect could be used to constrain supernova explosion models in the
future.  For the published refereed journal article that first pointed
out the significance of internal conversion electrons on supernova
light curves, please see \cite{seitenzahl2009d}.

\section{Nuclear decays}
\label{sec:decay}
The time dependence of $n$ nuclide abundances $N_i$ in a ecay
chain is governed by the Bateman equations:
\begin{eqnarray}
  \label{eq:bateman}
  \frac{dN_1}{dt}&=&-\lambda_1 N_1\\
  \frac{dN_i}{dt}&=& \lambda_{i-1} N_{i-1}-\lambda_{i} N_{i} \hspace{8pt}.
\end{eqnarray}
For $n=2$ and initial abundances $N_1(0)$ and $N_2(0)$ we get the
solution
\begin{eqnarray}
  \label{eq:sol1}
  N_1(t)&=& N_1(0)\exp(-\lambda_1 t) \\
  \label{eq:sol2}
  N_2(t)&=& N_1(0) \frac{\lambda_1}{\lambda_2-\lambda_1} [
  \exp(-\lambda_1 t)-\exp(-\lambda_2 t) ] + N_2(0)\exp(-\lambda_2 t) \hspace{8pt}.
\end{eqnarray}
The decay constants $\lambda_i$ are related to the half-lives
$t^{1/2}_i$ and the mean life-time $\tau_i$ via
\begin{equation}
  \lambda_i = \frac{1}{\tau_i} = \frac{\ln(2)}{t^{1/2}_i} \hspace{8pt}.
\end{equation}
The rate of energy deposition by decays of nucleus $i$ is given by the
activity multiplied by the energy deposited per decay:
\begin{equation}
  \epsilon_i = \lambda_i N_i(t) q_i(t) \hspace{8pt},
\end{equation}
where the number $N_i$ is given by eq.~\ref{eq:sol1} or
eq.~\ref{eq:sol2} and the energy deposited, $q_i$, is a function of
time due to the increasing escape fraction of $\gamma$-rays and
possible late time escape of positrons.
\vspace{12pt}
\begin{wraptable}[8]{}{9cm}
\small
\vspace{-24pt}
    \caption{Radioactive decay energies \label{tab1} (keV
      decay$^{-1}$)} \vspace{5pt}
    \begin{tabular}{ccccc} \hline {Nucleus} & {Auger $e^-$ } & {IC
        $e^-$} & {$e^+$} & {X-ray } \\ \hline
      \nuc{57}{Co} &7.594&10.22&0.000&3.598\\
      \nuc{56}{Co} &3.355&0.374&115.7&1.588\\
      \nuc{55}{Fe} &3.973&0.000&0.000&1.635\\
      \nuc{44}{Ti} &3.519&7.064&0.000&0.768\\
      \nuc{44}{Sc} &0.163&0.074&595.8&0.030\\ \hline
    \end{tabular}
\end{wraptable}
Most of the nuclei synthesized in supernova explosions have atomic
masses in the range $A\approx 12-70$.  The isotopes are generally
either stable or on the proton-rich side of the valley of stability.
For these unstable nuclei, radioactive decay occurs along an isobar
towards neutron richness, either by electron capture or positron
emission.  Electron capture proceeds via the capture of an atomic
(typically K or L shell) electron by a proton in the nucleus and
corresponding emission of an electron neutrino.  Positron emission
proceeds via the decay of a proton in the nucleus into a neutron and
the corresponding emission of a positron and an electron neutrino.
For both processes, the transition to the daughter nucleus is a
statistical process to a distribution of (excited) nuclear
states. Following electron capture, the daughter is formed with a hole
in its atomic electron cloud.  Higher lying atomic electrons
transition to fill the gaps in the lower lying atomic shells, which
results in characteristic X-rays being emitted from the cascade.  For
every such electron transition, there is also a probability that
instead of an X-ray one or more higher lying atomic electrons are
ejected. These electrons are known as Auger electrons.
\begin{wrapfigure}[16]{R}{2.5in}
  \includegraphics{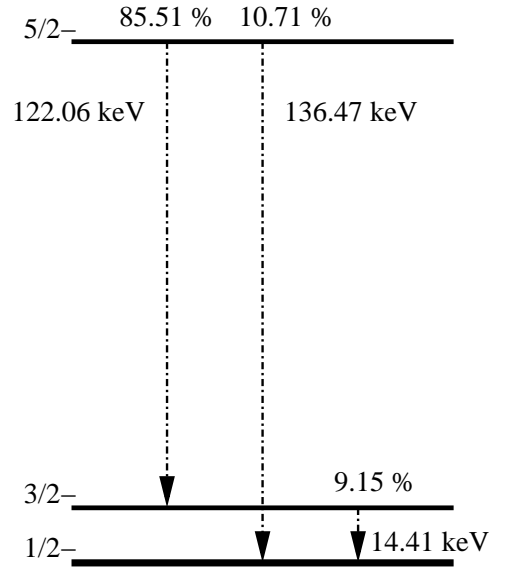}
  \caption{\label{fig1} Ground state and first two exited levels of
    \nuc{57}{Fe} showing $\gamma$-ray energies and intensities for the
    decay of \nuc{57}{Co}.}
\end{wrapfigure}
Analogous to the X-ray cascade of the atom, the excited daughter
nucleus typically undergoes a series of transitions towards the ground
state accompanied by emission of characteristic $\gamma$-ray photons.
If the nucleus is surrounded by an electron cloud, then there is again
a probability that instead of a $\gamma$-ray one or more atomic
electrons are ejected. This process, in which the energy difference of
the nuclear levels is carried away by the ejection of an inner atomic
electron and there is no $\gamma$-ray photon emitted, is called
internal conversion.  The probability for internal conversion to occur
for a given level is measured by the internal conversion coefficient
$\alpha = \frac{\# \mathrm{of} \; e^- \; \mathrm{de-excitations}}{\#
  \mathrm{of} \; \gamma \; \mathrm{de-excitations}}$. $\alpha$ is
normally small for nuclei with low atomic number, but increases for
levels close to the ground state.  For the decay of \nuc{57}{Co} the
probability for production of internal conversion electrons is very
large, due to the existence of a low-lying nuclear level in the
daughter nucleus \nuc{57}{Fe} (see Fig.~\ref{fig1}).  For the first
exited state (14.4 keV $3/2^{-}$) of \nuc{57}{Fe}, the internal
conversion coefficient $\alpha = 8.58$.  This decay is 100\% electron
capture, and over 99.8\% all decays are into the 136 keV
level. Captures to the higher lying 366.74 and 706.42 keV levels
contribute only marginally.  For ground-state to ground-state electron
capture transitions, such as in the decay of \nuc{55}{Fe}, no
$\gamma$-rays or positrons are emitted.  Assuming that the neutrino
escapes without interactions, in such transitions the Auger electrons
and the X-rays constitute the only sources of radioactive heating.

\noindent The following four decay chains contribute most to
bolometric supernova light curves:
\begin{align}
  & &^{56}\mathrm{Ni} \;\stackrel{t_{1/2} = \; 6.08d}{\hbox to
    60pt{\rightarrowfill}} \; ^{56}\mathrm{Co} \;
  \stackrel{t_{1/2} = \; 77.2d}{\hbox to 60pt{\rightarrowfill}} \; ^{56}\mathrm{Fe} \\
  & &^{57}\mathrm{Ni} \;\stackrel{t_{1/2} = \; 35.60 h}{\hbox to
    60pt{\rightarrowfill}}\; ^{57}\mathrm{Co} \;
  \stackrel{t_{1/2} = \; 271.79d}{\hbox to 60pt{\rightarrowfill}} \; ^{57}\mathrm{Fe}\\
  & &^{55}\mathrm{Co} \;\stackrel{t_{1/2} = \; 17.53 h}{\hbox to
    60pt{\rightarrowfill}}\; ^{55}\mathrm{Fe} \;
  \stackrel{t_{1/2} = \; 999.67 d}{\hbox to 60pt{\rightarrowfill}} \; ^{55}\mathrm{Mn}\\
  & &^{44}\mathrm{Ti} \;\stackrel{t_{1/2} = \; 58.9 y}{\hbox to
    60pt{\rightarrowfill}}\; ^{44}\mathrm{Sc} \; \stackrel{t_{1/2} =
    \; 3.97 h}{\hbox to 60pt{\rightarrowfill}} \; ^{44}\mathrm{Ca}
\end{align}
Here we do not model the radiative transport, we only compare leptonic
(and X-ray) energy injection rates.  We do not include the energy
produced by the pair annihilation and further assume that the kinetic
energy of the leptons is completely and locally deposited and
thermalized. The energy generation rates presented here do not include
any heating due to $\gamma$-rays and therefore are not predictions for
bolometric light curves.  This approach, however, does allow for a
direct comparison of the relative importance of the different leptonic
heating channels (positrons from the decays of \nuc{56}{Co} and
\nuc{44}{Sc} and electrons from the decays of \nuc{57}{Co} and
\nuc{55}{Fe}.)  The relevant energies of the different decay channels
are listed in table~\ref{tab1}.  These data are taken from the
National Nuclear Data Center\footnote{http://www.nndc.bnl.gov/}.
\section{Late-time bolometric light curves}
\label{sec:lc}
Bolometric light curves have to be reconstructed from multi-band
photometry.  For reliable reconstruction, the contribution of the
UV/optical ($UBV\!RI$) and near-infrared ($JHK$) bands have to be
included (UVOIR light curve). However, near-IR observations are rare
at very late epochs, and sometimes only $B$-through-$I$ band
observations with a near-IR correction extrapolated from earlier
epochs are used \cite{sollerman2002a}. Below, we discuss some aspects
and examples of different supernova light curves. 

\subsection{\it Type Ia supernovae \label{sec:snia}}
\begin{SCfigure}[1][ht!]
    \includegraphics[width=8cm,clip]{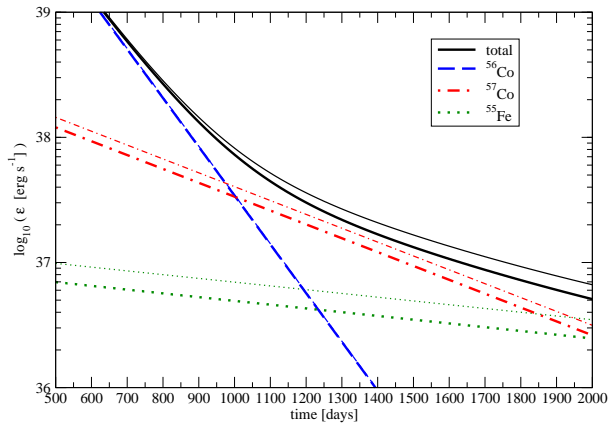}
    \caption{\label{fig2} Instantaneous energy generation rates for
      initial abundances taken from the W7 model of
      \cite{iwamoto1999a}.  Thick lines are due to positrons, IC and
      Auger electrons alone. The thin lines also include the full
      X-ray dose, which is small even in the limit of complete X-ray
      trapping.  The contribution of electrons and positrons from the
      decays of \nuc{44}{Ti} and \nuc{44}{Sc} is too small to be seen
      in this figure.}
\end{SCfigure}
\noindent SNe~Ia are thought to be thermonuclear disruptions of white dwarf
stars and as such do not have an extended envelope. Consequently,
their ejecta become transparent to $\gamma$-rays relatively soon and
the positron-dominated phase generally starts $150$--$300$ days after
the explosion \cite{milne2001a,sollerman2004a}.  Between $300$ and
$600$ days, the UVOIR light curves fall exponentially with the
\nuc{56}{Co} decay half-life \cite{stritzinger2007a}, a clear sign
that a constant fraction of positrons is trapped.  Unfortunately, only
few SN~Ia were observed at even later times, but there are indications
of a slow-down in the light curves of some optical bands after
$\sim$\,$600$ days \cite{sollerman2004a,lair2006a}.  A particularly
interesting case is SN~2003hv, which shows a slow-down in the {\em
  bolometric} light curve 786 days after B-band maximum light
\cite{leloudas2009a}.  It has been suggested that additional (possibly
radioactive) heating sources may be needed to explain this observed
effect.  We point out that the magnitude and time of occurrence of the
observed slow down of the light curve is expected and a natural
consequence of the leptonic energy injection from the decay of
\nuc{57}{Co}.  To illustrate this, consider the leptonic energy
generation rates of important long lived isotopes between 500 and 2000
days for W7 \cite{nomoto1984a}, a common reference SN~Ia model (see
Fig.~\ref{fig2}).  For this particular choice of yields
\cite{iwamoto1999a}, the light curve is significantly higher from
assuming only heating from \nuc{56}{Co} positrons after about 750
days.  1000 days after the explosion, the heating from \nuc{57}{Co}
electrons equals the heating from \nuc{56}{Co} positrons.

\subsection{\it Core collapse supernovae\label{sec:ccsn}}
The progenitors of core collapse SNe are massive stars, which have an
extended, massive envelope at the time of explosion, which provides
opacity for the emerging $\gamma$-rays. Compared to SNe~Ia, the
nuclear burning also takes place at higher entropy, which means that
core collapse supernovae synthesize more \nuc{44}{Ti}. These
differences imply that the moment when internal conversion electrons
from \nuc{57}{Co} significantly contribute to the light curves is
delayed. Nevertheless, two observed examples of core collapse SN where
internal conversion electrons do significantly alter the shape of the
light curve are given below.

\subsubsection{SN 1998bw}
\begin{SCfigure}[1][ht!]
  \includegraphics[width=8cm,clip]{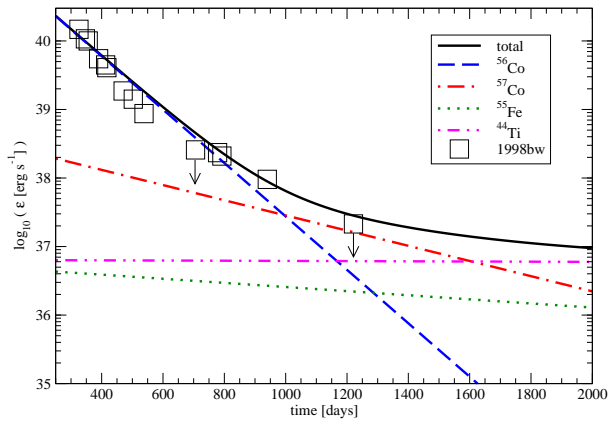}
  \caption{\label{fig3} Instantaneous energy generation rates due to
    radioactivity for initial abundances taken from a hypernova model
    of \cite{nakamura2001a}. Only leptons and X-rays (which are small
    compared to the leptonic contribution) are included (i.e. no
    $\gamma$-ray contribution or freeze-out effects are
    considered). Squares are the data of 1998bw of
    \cite{sollerman2002a}. Arrows represent $3\sigma$ upper limits.
    There are no free parameters. }
\end{SCfigure}
\noindent The well observed SN~1998bw, which was the first case of a
$\gamma$-ray burst associated with a SN\cite{galama1998a}, was
classified as a SN~Ic. The explosion was a very energetic , asymmetric
disruption of a massive, stripped stellar core ($\geq 10$ \msun;
\cite{maeda2002a}). The high ejecta velocities and the lack of an
extended envelope reduce the $\gamma$-ray opacity such that by $1000$
days almost all $\gamma$-rays are freely streaming.  A nucleosynthetic
yield calculation for hypernovae like 1998bw \cite{nakamura2001b}
predicts a \nuc{57}{Ni} to \nuc{56}{Ni} mass ratio
$\mathcal{R}^{57/56} \approx 0.0366$, $\sim 1.5$ times the solar value
for the \nuc{57}{Fe} to \nuc{56}{Fe} ratio (which is $0.0234$;
\cite{lodders2003a}).  After modeling the UVOIR light curve to
$\sim$\,$1000$ days after the explosion, \cite{sollerman2002a} showed
that a simple model without contributions from freeze-out effects,
circumstellar interaction, accretion onto a central compact object or
light echoes requires a \nuc{57}{Ni} to \nuc{56}{Ni} ratio $\sim 13.5$
times greater than solar.  This large discrepancy between the value of
$\mathcal{R}^{57/56}$ predicted from explosive nucleosynthesis
calculations and the one derived from light curve modeling reduces if
the effects of internal conversion and Auger electrons are included in
the light-curve calculations.  In Fig.~\ref{fig3} we show the combined
leptonic and X-ray luminosity corresponding to the nucleosynthetic
yield calculations of \cite{nakamura2001b}.  The non-$\gamma$-ray
heating due to \nuc{57}{Co} is the dominant contribution between
$\sim1000$ and 1600 days.  The observed slow down of the light curve
of 1998bw at $\sim900$ days (see Fig.~\ref{fig3}) is thereby naturally
obtained without the need for strongly super solar
$\mathcal{R}^{57/56}$.

\subsubsection{SN 1987A}
\begin{SCfigure}[1][ht!]
    \includegraphics[width=8cm,clip]{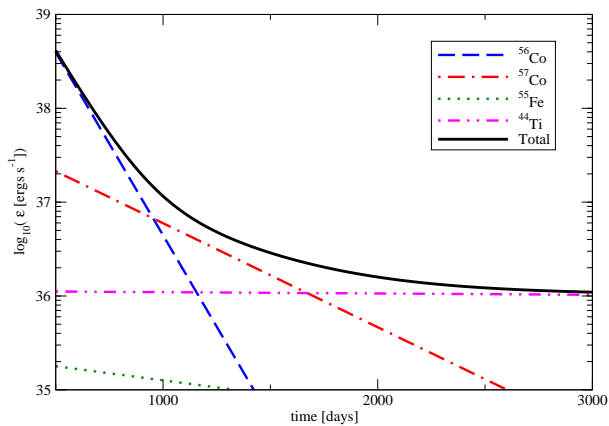}
    \caption{Leptonic energy deposition rates for important isotopes
      of the predicted yields for 1987A of \cite{nomoto1988a}. Note
      the hitherto ignored leptonic contribution of \nuc{57}{Co} (dot
      dashed) and its impact on the leptonic heating rate between 1000
      and 2000 days.}
    \label{fig1987a}
\end{SCfigure}
\noindent SN~1987A has demonstrated that light curves can be constructed for
nearby supernovae extending for several years after the
explosion. This has led to the consideration of longer lived
radioactive species.  The most important of these is \nuc{57}{Co},
which is expected to be produced in significant amounts as the decay
product of the short lived \nuc{57}{Ni}.  Due to its relatively longer
half-life (271.79 days) and the high opacity of the ejecta to the
emitted $\gamma$-rays, its $\gamma$-rays alone may dominate the
bolometric light curve, especially of core collapse supernovae, at
late times \cite{pinto1988a,woosley1989a}. In fact, due to its
extended envelope, the opacity to $\gamma$-rays in the remnant of
1987A remained high for such a long time that it never entered the
\nuc{56}{Co} positron dominated phase \cite{fransson2002a}.  It is
argued, that the only other radionuclide which noticeably contributes
to the bolometric light curve is \nuc{44}{Ti}
\cite{kumagai1989a}. \nuc{44}{Ti} has an even longer half-life of 59.8
years, and the short lived daughter \nuc{44}{Sc} has a strong positron
channel, which means that the \nuc{44}{Ti} decay chain is the dominant
energy source at very late times.  Photometric data of SN 1987A lead
to the conclusion that the $\mathcal{R}^{57/56}$ was rather large,
about five times the corresponding solar value
\cite{suntzeff1992a,dwek1992a}.  Due to the proximity of SN~1987A, the
presence of \nuc{56}{Co} and \nuc{57}{Co} was inferred not only
photometrically from their imprint on the light curve, but also
through a direct detection of escaping $\gamma$-rays from their
decays.  The observed gamma ray flux from \nuc{57}{Co} with the
Oriented Scintillation Spectrometer Experiment (OSSE) on board the
{\em Compton Gamma Ray Observatory} \cite{kurfess1992a} favored at
most moderate enhancement of \nuc{57}{Ni} (factor 1.5), in good
agreement with nuclear reaction network calculations
\cite{nomoto1988a}. Similarly, based on the light curve, a rather high
\nuc{44}{Ti} abundance was claimed in several studies
\cite{dwek1992a,woosley1991a, timmes1996a}.  The great difficulty of
explaining the light curve of 1987A between 1000 and 1500 days without
strong overproduction of \nuc{57}{Ni} can be ameliorated by including
``freeze-out effects'' \cite{fransson1993a} or by conveniently
re-arranging the spatial distribution of the radionuclides
\cite{clayton1992a}, but some disagreement of data and models
persists.
By comparing the various leptonic energy deposition rates for a
nucleosynthesis model of 1987A \cite{nomoto1988a}, we show
that the leptonic channels of \nuc{57}{Co} decay are significantly
contributing between 1000 and 2000 days (see Fig.~\ref{fig1987a}).

\section{Conclusion and Outlook}
\label{sec:conc}
Fitting models to observed late time light curves provides a unique
and independent method to directly measure the {\it isotopic} yields
of prominent radioactive nuclei synthesized in the explosion (in
particular \nuc{56}{Ni}, \nuc{57}{Ni}, and \nuc{44}{Ti}).  We have
shown that at late times, when the ejecta have become largely
transparent to $\gamma$-rays, the energy carried by Auger and internal
conversion electrons may constitute a significant source of
heating. These additional decay channels have to be considered for
reliable isotopic abundance determinations from light curves.  In
particular, we have shown that a re-analysis of the bolometric light
curve of 1987A (taking the hitherto unconsidered effect of internal
conversion electrons into
account) would likely yield significantly different (smaller)
\nuc{44}{Ti} and \nuc{57}{Ni} masses.  The new derived values for
\nuc{57}{Ni} and \nuc{44}{Ti} would allow us to gain more insight into
the explosion mechanism of core collapse supernovae. In particular,
observationally driven inferences about the location of the mass cut
can be made, which would give us valuable information about the
physical processes separating neutron star and black hole formation
from massive stars.  Last but not least, SNe~Ia are considered as the
source of the positrons required to explain the Galactic 511 keV
annihilation line observed by Integral/SPI
\cite{knoedlseder2005a}. The question whether enough positrons can
escape the remnant remains unanswered.  An understanding of the
physics that underlies the light curves of SNe is a crucial step in
order to constrain the escape fraction of positrons.


\end{document}